\documentclass[12pt]{article}\usepackage{Sep_CMI}

\begin{document}

\title{Separability of Density Matrices\\
 and\\
Conditional Information Transmission}

\author{Robert R. Tucci\\
        P.O. Box 226\\ 
        Bedford,  MA   01730\\
        tucci@ar-tiste.com}

\date{ \today} 

\maketitle

\vskip2cm
\section*{Abstract}
We give necessary and sufficient conditions under which a density matrix acting on a two-fold tensor
product space is separable. Our conditions are given in terms of quantum conditional information transmission.

\newpage
Ref.\cite{Tucci-tang99}
proposed 
using 
quantum conditional information transmission
as a measure of entanglement. In its simplest case, 
this measure requires
one speaker and two listeners. On the other hand,
the simplest case of separability of 
density matrices is defined for two listeners but no speaker. 
Thus, it is not immediately apparent how
quantum conditional information transmission
is related to separability. And yet, they 
must be closely related since they are
both closely related to the phenomenon of 
quantum entanglement.
In this paper, we present a theorem that elucidates the hidden relationship
between conditional information transmission
and separability. 
The theorem gives necessary and sufficient conditions for the separability
of density matrices acting on a two-fold tensor product space.
The theorem can be easily generalized to the case of $n$-fold tensor products.

We will use $\hil_\rva, \hil_\rvb, \ldots$ to represent   Hilbert spaces 
(finite dimensional ones for simplicity), 
and $\hil_{\rva, \rvb}$ to represent $\hil_\rva \otimes \hil_\rvb$, the tensor product of 
$\hil_\rva$ and $\hil_\rvb$. $dim(\hil)$ will stand for the dimension of 
the Hilbert space $\hil$. The set of all density matrices acting on a Hilbert space $\hil$
will be denoted by $\den(\hil)$. If $\rhoxy \in \den(\hilxy)$, we will denote
the partial traces of $\rhoxy$ by $\rhox = \tr_\rvy \rhoxy$ and $\rhoy = \tr_\rvx \rhoxy$.
For any set $S$, we will use $|S|$ to represent the number of elements in $S$.

We will say  $\rho \in \den(\hilxy)$ is {\it separable} (or more precisely,  $\rvx, \rvy$ {\it separable})
 if $\rho$ can be expressed as

\beq
\rho = \sum_e w_e \rhox^e  \rhoy^e
\;,
\label{eq:xysep}\eeq
where the $w_e$, called {\it weights}, are non-negative numbers that sum to 1, and where for all $e$,
$\rhox^e \in \den(\hilx)$ and  $\rhoy^e \in \den(\hily)$.
{\it Non-entangled} $\rvx, \rvy$ states are usually defined as those 
which are $\rvx,\rvy$ separable.

We will say  $\rho \in \den(\hilxye)$ is {\it conditionally separable} (or more precisely,
$\rvx, \rvy| \rve$ {\it separable})
if $\rho$ can be expressed as

\beq
\rho = \sum_e w_e \ket{e} \bra{e} \rhox^e  \rhoy^e
\;,
\label{eq:xyesep}\eeq
where the $w_e$, called {\it weights}, are non-negative numbers that sum to 1, the states $\ket{e}$ 
are an orthonormal basis for $\hil_\rve$, 
and for all $e$,
$\rhox^e \in \den(\hilx)$ and  $\rhoy^e \in \den(\hily)$.

Suppose $A$ is a set of random variables. For example, $A =\{ \rva, \rvb\}$. If $\rho\in \den(\hil_A)$ 
and $A'\subset A$, then we will use $S_\rho(A')$ to represent $S(\tr_{A-A'} \rho)$,
where $S(\cdot)$ is the von Neumann entropy. For example,
if $\rho\in \den(\hil_{\rva, \rvb})$, then $S_\rho(\rva) = S(\tr_\rvb \rho)$. If $\rho \in \den(\hilxye)$,
we define the quantum {\it conditional mutual information}, or {\it conditional information transmission} by

\beq
S_\rho[ (\rvx: \rvy) | \rve] =
S_\rho(\rvx, \rve)  + S_\rho(\rvy, \rve) - S_\rho(\rvx, \rvy, \rve) - S_\rho(\rve)
\;.
\eeq
The classical counterpart of this is the classical conditional mutual information, 
which is defined, for random variables $\rvx, \rvy, \rve$ with a
joint distribution $P(x,y,e)$, by

\beq
H[ (\rvx: \rvy) | \rve] = \sum_{x,y,e} P(x,y,e) \log_2 \frac{ P(x,y|e)}{P(x|e)P(y|e)}
\;.
\eeq
See Ref.\cite{Tucci-review}
for a review of classical and quantum entropy
presented in the same notation used in this paper.

For any $\rho\in \den(\hilxye)$, 

\beq
S_\rho[(\rvx : \rvy) | \rve] \geq 0
\;.
\label{eq:q-ssa-ineq}\eeq
This is called the {\it strong subadditivity 
inequality} for quantum entropy. It was first proven by Lieb-Ruskai in Ref.\cite{LR}.
More recently, it has been shown\cite{Petz-LR} that the strong subadditivity inequality
becomes an equality (i.e., ``is saturated") if and only if $\rho$ satisfies

\beq
\log \rho = \log \rhoxe + \log \rhoye - \log \rhoe
\;.
\label{eq:q-log-cond-ind}\eeq
Classical random variables $\rvx, \rvy, \rve$ with joint distribution $P(x,y,e)$
satisfy

\beq
H[(\rvx : \rvy) | \rve] \geq 0
\;,
\label{eq:c-ssa-ineq}\eeq
which is the classical counterpart of Eq.(\ref{eq:q-ssa-ineq}).
This inequality is saturated if and only if

\beq
P(x, y|e) = P(x|e)P(y|e)
\;
\label{eq:c-cond-ind}\eeq
for all $x,y,e$. When Eq.(\ref{eq:c-cond-ind}) is true, we say
$\rvx, \rvy$ are {\it conditionally independent}.
Taking the logarithm of both sides of Eq.(\ref{eq:c-cond-ind}) yields 

\beq
\log P(x, y, e) = \log P(x, e) + \log P(y, e)  - \log P(e)
\;,
\eeq
which is the classical counterpart of Eq.(\ref{eq:q-log-cond-ind}).

\vspace{3ex}
\noindent
{\bf Theorem 1:} 
$\rho\in \den(\hilxy)$ is $\rvx, \rvy$ separable
if and only if
there exists a Hilbert space $\hile$ and a density matrix $\sigma \in \den(\hilxye)$ such that
\begin{enumerate}
\item $\rho = \tr_\rve \sigma$,
\item $S_\sigma[(\rvx : \rvy) | \rve] = 0$,
\item $\sigye, \sigxe$ and $\sige$ commute pairwise,
\item the eigenvalues of $\sige$ are are non-zero and non-degenerate.
\end{enumerate}
\noindent
{\it proof:} 

($\Rightarrow$) $\rho$ can be expanded as in Eq.(\ref{eq:xysep}).
We can always choose the weights $w_e$ of the expansion to be non-zero and non-degenerate. 
Indeed, if $w_e = 0 $, we just eliminate that term from the expansion.
If $e_1 \neq e_2$ and $w_{e_1} = w_{e_2}$, then we replace
the $e_1$ and $e_2$ terms of the expansion by

\begin{eqnarray}
\lefteqn{w_{e_1} (\rho_\rvx^{e_1} \rhoy^{e_1} + \rhox^{e_2} \rhoy^{e_2} ) =} \nonumber \\
&& =w_{e_1} \rhox^{e_1} \rhoy^{e_1}
+ (\frac{w_{e_1}}{2} + \epsilon)\rhox^{e_2} \rhoy^{e_2} 
+ (\frac{w_{e_1}}{2} - \epsilon)\rhox^{e_2} \rhoy^{e_2} = \nonumber \\
&&=\sum_{j = 1}^3 w'_{e_j} \rhox^{'e_j} \rhoy^{'e_j}
\;,
\end{eqnarray}
where we have define a new $e$ value called $e_3$ and 
we have set $w'_{e_1} = w_{e_1}$, $w'_{e_2} = \frac{w_{e_1}}{2} + \epsilon$ 
and  $w'_{e_3} = \frac{w_{e_1}}{2} - \epsilon$. For small enough $\epsilon>0$,
we achieve our goal of representing $\rho$ as in Eq.(\ref{eq:xysep}) with
weights that are non-degenerate and non-zero. If
$E$ is the new set of $e$ values,
let $\hil_\rve$ be a Hilbert space of dimension $|E|$,
and let $\ket{e}$ for $e\in E$ be an orthonormal basis for $\hil_\rve$.
Define $\sigma \in \den(\hilxye)$ by

\beq
\sigma = \sum_{e\in E} w_e \ket{e} \bra{e} \rhox^e  \rhoy^e
\;.
\eeq
Thus, $\sigma$ is $\rvx, \rvy | \rve$ separable.
Clearly, $\rho = \tr_\rve \sigma$.
In Ref.\cite{Tucci-tang99}, it is shown by straightforward computation 
that any $\rvx, \rvy | \rve$ separable density matrix $\sigma$ 
satisfies $S_\sigma[(\rvx : \rvy) | \rve] = 0$.
$\sigma$ has the following partial traces:

\beq
\sige = \sum_e w_e \ket{e}\bra{e}
\;,
\eeq

\beq
\sigxe = \sum_e w_e \ket{e}\bra{e} \rhox^e
\;,
\eeq

\beq
\sigye = \sum_e w_e \ket{e}\bra{e} \rhoy^e
\;.
\eeq
Clearly, $\sigye, \sigxe$ and $\sige$ commute pairwise.
The eigenvalues of $\sige$ are the $w_e$, which are non-zero and non-degenerate.

($\Leftarrow$)
$S_\sigma[(\rvx : \rvy) | \rve] = 0$ so Eq.(\ref{eq:q-log-cond-ind}) is true for $\sigma$.
In fact, since $\sigye, \sigxe$ and $\sige$ commute pairwise, and $\rhoe$ has non-zero
eigenvalues, we can combine the logarithms to obtain

\beq
\sigma = \sigye \sigxe (\sige)^{-1}
\;.
\label{eq:sig-div}\eeq
Since $\sige$ is a Hermitian matrix, it can be diagonalized:

\beq
\sige = \sum_e w_e \ket{e} \bra{e}
\;,
\eeq
where $w_e$ and $\ket{e}$ for all $e$ are the eigenvalues and eigenvectors of $\sige$.
One has that

\beq
\sige \sigxe = \sigxe \sige
\;.
\eeq
Thus,

\beq
w_e \bra{e}\sigxe \ket{e'}= \bra{e}\sigxe \ket{e'} w_{e'}
\;
\eeq
for all $e,e'$. Since the eigenvalues $w_e$ of $\sige$ are non-degenerate,
$e\neq e'$ implies  $w_e \neq w_{e'}$, and therefore $\bra{e}\sigxe \ket{e'} = 0$.
It follows that $\sigxe$ is diagonal in its $\hile$ sector:
\beq
\sigxe = \sum_{x,x',e} A^e_{x,x'} \ket{e,x}\bra{e,x'}
\;,
\label{eq:sigxe1}\eeq
where for all $x,x',e$, $A^e_{x,x'}$ is a complex number, and where $\ket{x}$ for all $x$ is any orthonormal basis of $\hilx$.
If for each $e$, $\rhox^e \in \den(\hilx)$ is defined by

\beq
\rhox^e = \sum_{x,x'} \frac{A^e_{x,x'}}{w_e} \ket{x}\bra{x'}
\;,
\eeq
then 
Eq.(\ref{eq:sigxe1}) can be rewritten as

\beq
\sigxe = \sum_e w_e \ket{e}\bra{e} \rhox^e
\;.
\label{eq:sigxe2}\eeq
By a similar argument, $\sigye$ is also diagonal in its $\hile$ sector and can be expressed as 

\beq
\sigye = \sum_e w_e \ket{e}\bra{e} \rhoy^e
\;,
\label{eq:sigye2}\eeq
where for all $e$, $\rhoy^e \in \den(\hily)$.
Our newly found, diagonal in the $\hile$ sector,
expressions for $\sigye, \sigxe$ and $\sige$  can now be substituted
into  Eq.(\ref{eq:sig-div})  to get 

\beq
\sigma = \sum_e w_e \ket{e}\bra{e} \rhox^e \rhoy^e
\;.
\eeq
Thus, $\sigma$ is $\rvx, \rvy| \rve$ separable. 
Taking the $\rve$ trace of this $\sigma$ to get $\rho$,
we see that $\rho$ is $\rvx, \rvy$ separable. QED 

There probably exist certain $\rho\in \den(\hilxy)$ for which  conditions 1 to 4 on the right hand side of Theorem 1
cannot be achieved for finite $dim(\hil_\rve)$, but can be achieved in
the limit $dim(\hil_\rve)\rarrow \infty$. Such $\rho$ could be described
as being weakly separable.

Let $\den_{insep}(\hilxy)$ be the set of  all 
$\rho \in \den(\hilxy)$ which are not $\rvx, \rvy$ separable. Let
$\den_{pos}(\hilxy)$ be the set of  all $\rho \in \den(\hilxy)$
for which all extensions $\sigma \in \den(\hilxye)$ such that $\rho = \tr_\rve\sigma$
satisfy $S_\sigma[(\rvx : \rvy) | \rve] \neq 0$. Then,
by Theorem 1, $\den_{pos}(\hilxy) \subset \den_{insep}(\hilxy)$.
Density matrices in $\den_{pos}(\hilxy)$ and those in  $\den_{insep}(\hilxy) - \den_{pos}(\hilxy)$ 
exhibit different kinds of entanglement.

Some goals for future research are: give concrete examples of 
Theorem 1; explore the connection between Theorem 1 and   
the necessary condition for separability given by Peres\cite{Peres}, 
and the bound entanglement discovered by Horodecki.\cite{bound-entan}.

\vspace{3ex}
\noindent
{\it Acknowledgements:} I thank  M.A. Nielsen, D. Petz and M.B Ruskai 
for their generosity in communicating to me that the quantum strong subadditivity
inequality is saturated iff Eq.(\ref{eq:q-log-cond-ind}).


\begin{thebibliography}{99}
\bibitem{Tucci-tang99}
R.R. Tucci, ``Quantum Entanglement and Conditional Information Transmission", Los Alamos eprint quant-ph/9909040 .

\bibitem{Tucci-review}
R.R. Tucci, ``Quantum Information Theory - A Quantum Bayesian Nets Perspective", Los Alamos eprint quant-ph/9909039 .

\bibitem{LR}
E. Lieb, M.B. Ruskai, J. of Math. Phys. {\bf 14}, 1938 (1973).

\bibitem{Petz-LR}
This result was communicated to the author by 
M.B. Ruskai and M.A. Nielsen, who intend to publish their own proof in the near future.
The result was also independently communicated to the author by D. Petz, who pointed out
that strong subadditivity follows from  monotonicity, so this result is a corollary
of his paper: Commun. Math Phys. {\bf 105}, 123 (1986). See also pages 23-24 of
M. Ohya, D. Petz, {\it Quantum Entropy and Its Use} (Springer-Verlag, Heidelberg, 1993).

\bibitem{Peres}
A. Peres, Phys. Rev. Lett. A {\bf 76}, 1413 (1997). Also available as 
Los Alamos eprint quant-ph/9604005 .

\bibitem{bound-entan}
P. Horodecki, Phys. Rev. Lett. A {\bf 232}, 333 (1997). Also available as 
Los Alamos eprint quant-ph/9703004 .


\end{thebibliography}
\end{document}